\documentclass{PoS}
\usepackage{multirow}
\title{Charm physics with physical light and strange quarks using domain wall fermions }

\ShortTitle{Charm physics with physical ligth quarks}

\author{\speaker{Andreas J\"uttner}, Francesco Sanfilippo, Justus Tobias Tsang\\
        School of  Physics and Astronomy, University of Southampton, Highfield, SO17 1BJ Southampton, UK\\
        E-mail: \email{juettner@soton.ac.uk}, \email{fr.sanfilippo@gmail.com}, \email{jtt1g12@soton.ac.uk}}

\author{Peter~A.~Boyle, Luigi Del Debbio, Ava Khamseh\\
        School of Physics \& Astronomy, University of
        Edinburgh, EH9 3JZ, UK\\
        E-mail: \email{paboyle@ph.ed.ac.uk}, \email{luigi.del.debbio@ed.ac.uk}, \email{s0948358@sms.ed.ac.uk}}
\author{Nicolas~Garron\\
	Department of Applied Mathematics \& Theoretical Physics,
	University of Cambridge,
	Wilberforce Road, Cambridge CB3 0WA, United Kingdom 
        E-mail: \email{ng389@damtp.cam.ac.uk}}
 
\author{Marina Marinkovic\\
        School of  Physics and Astronomy, University of Southampton, Highfield, SO17 1BJ Southampton, UK and\\
        CERN, Physics Department, TH Unit, CH-1211 Geneva 23, Switzerland\\ 
        E-mail: \email{marina.marinkovic@cern.ch}}

\abstract{We present a study of charm physics using RBC/UKQCD's 
	2+1 flavour physical point Domain Wall Fermion ensembles for the light
        quarks as well as for the valence charm quark. 
	After a brief 
	motivation of domain wall fermions as a suitable heavy quark discretisation we will 
	show first results for masses and matrix elements.}

\FullConference{The 32nd International Symposium on Lattice Field Theory,\\
		23-28 June, 2014\\
		Columbia University New York, NY}

\begin{document}
\section{Introduction}
This talk presented our efforts towards a charmed meson physics program on RBC/UKQCD's
$N_f=2+1$ domain wall ensembles using the same discretisation for the light
and the charm quarks. 
Compared to lattice QCD for light quark physics, simulations for 
charm  are comparatively scarce (cf. the FLAG report~\cite{Aoki:2013ldr}). 
One of the reasons is that the charm quark sets a mass scale which until 
recently was very difficult  to reconcile  with the energy scale set by 
the light quarks in a fully
relativistic and dynamical lattice setup. In fact,
covering both energy scales at the same time still poses a challenge
due to the large lattice size that one needs to simulate in order to keep both
finite volume and cutoff effects reliably under control.

Our programme consists of first studying domain wall fermions as a charm quark
discretisation  within the quenched approximation. Let us emphasise 
that the intention is not to make
any phenomenologically relevant predictions from within this framework 
but rather to understand the behaviour of domain wall
fermions for heavy quarks and the continuum limit in detail. To this end
we created ensembles and measurements for mesonic matrix elements
for cutoffs in the range of 2-6GeV. 
While a similar study with dynamical domain wall quarks is still out of question for the
foreseeable future we expect a number of properties to be very similar.
In particular, we expect that
the continuum limit scaling observed in the quenched theory over a large range
of lattice spacings will be qualitatively the same also beyond the quenched approximation.
We can therefore use the scaling behaviour found in the quenched theory to 
constrain contributions to the scaling from higher orders in the lattice
spacing when analysing data in the dynamical case but on a smaller 
range of lattice cutoffs. Our exploratory quenched studies have been
covered in Tsang's and Cho's talks at this conference~\cite{Tsangtalk,Chotalk}. 
The main result
is the observation of a large region of heavy quark masses for which 
the automatic $O(a)$-improvement of domain wall quarks is maintained and 
spectral quantities and matrix elements  extrapolate 
to the continuum limit linearly in $a^2$. We found that
by tuning the domain wall height, a free parameter in this discretisation,
 a reduction in the size of cutoff effects can be 
achieved. The experience gained in the quenched case is now being applied 
to a charm phenomenology program using dynamical gauge field ensembles. This
talk reports on the status of these simulations.

\section{Our strategy}
\begin{table*}
\begin{center}
        \small
\begin{tabular}{l@{\hspace{3mm}}l@{\hspace{2mm}}c@{\hspace{2mm}}c@{\hspace{2mm}}c@{\hspace{2mm}}c|@{\hspace{2mm}}clccc@{\hspace{2mm}}c@{\hspace{2mm}}c@{\hspace{2mm}}c}
 \hline\hline &&&&&&&&&\\[-3.0ex]
 $\beta$ & $a\,/\mathrm{fm}$ & $L/a$ & $T/a$
 &  $m_\pi$/MeV&$m_\pi L$&
 $\beta$ & $a\,/\mathrm{fm}$ & $L/a$ & $T/a$
 &  $m_\pi$/MeV&$m_\pi L$\\ \hline&&&&&&&&&\\[-4mm]
 2.13 &0.11 &24 &64  &431 &5.8&2.25 &0.08 &32 &64  &360 &4.8\\ 
 2.13 &0.11 &24 &64  &341 &4.6&2.25 &0.08 &32 &64  &304 &4.1\\
 2.13 &0.11 &48 &96  &139 &3.8&2.25 &0.08 &64&128  &139 &3.9\\
 \hline\hline
\end{tabular}
\end{center}
\caption{Parameters for ensembles processed so far: $\beta$ is the gauge
        coupling, $a$ is the lattice spacing, $L$ and $T$ are 
	the spatial and time-extent of the lattice, respectively,
        $m_\pi$ is the pion mass.}
\label{tab:ensemble-params}
\end{table*}
Table~\ref{tab:ensemble-params} gives an overview of RBC/UKQCD's domain wall ensembles
which we are currently analysing for this charm study. The ensembles to the left
have a lattice cutoff of 1.7-1.8GeV (coarse) and the ones to the right 2.4GeV (medium). 
We are currently generating an additional ensemble aiming at a third lattice spacing with 
$a^{-1}\approx 2.8$GeV and with a pion mass around 200MeV (fine). 
The global strategy can be summarised as follows: we will compute 
heavy-light, heavy-strange and heavy-heavy meson observables 
(masses, leptonic and semileptonic decay matrix elements, bag parameters)
on all ensembles  and
also the vector two-point function relevant for computing the anomalous magnetic moment
of the muon. Based on the findings in the quenched study we are certain that 
predictions for these quantities for the physical charm quark mass and even heavier,
towards the bottom quark mass, can be
made on the medium and fine ensembles with small cutoff effects and functioning
automatic $O(a)$-improvement. This may not be the case on the coarse ensemble where
the range in quark masses where we expect to have a good control over cutoff effects doesn't
allow to simulate directly at charm. On the coarse ensemble the bare input quark mass
corresponding to charm is above 0.45 and hence in the part of parameter space where
our numerical evidence from the quenched theory suggests the beakdown of 
$O(a)$-improvement~\cite{Tsangtalk}.
We therefore consider a strategy
where results for all quantities with heavy quark masses slightly smaller than
charm are made in the continuum limit. This is then followed by a small extrapolation in the
heavy quark mass towards the physical mass of charm. This does not preclude 
considering a global analysis ansatz taking advantage of the results for heavyier quark 
masses on the medium and fine ensemble once the latter has been generated and analysed.

In the following we present first results for the heavy-light and heavy-strange
decay constant. 

\section{Choice of parameters}
Our gauge field ensembles represent QCD with $N_f=2+1$
dynamical flavours at two different lattice 
spacings~\cite{Allton:2008pn,Aoki:2010pe,Aoki:2010dy,Blum:2014tka}.  The basic parameters of these
ensembles are listed in Table~\ref{tab:ensemble-params}.
For the lattices with physical light quark masses
we use the Iwasaki gauge action~\cite{Iwasaki:1984cj,Iwasaki:1985we}
and the domain wall fermion  action~\cite{Kaplan:1992bt,Shamir:1993zy}
with the M\"obius-kernel~\cite{Brower:2004xi,Brower:2005qw,Brower:2012vk}.
For the unphysical quark mass ensembles we use the \emph{standard} Shamir-kernel.
The difference between both kernels in our implementation~\cite{Blum:2014tka}
is a rescaling such that M\"obius domain wall fermions
are equivalent to Shamir domain wall fermions at twice the extension in the fifth
dimesion. M\"obius domain wall fermions are hence cheaper to simulate while providing
the same level of lattice chiral symmetry. Results from both formulations of domain wall
fermions lie on the same scaling trajectory towards the continuum limit where
cutoff-effects are of $O(a^2)$.
All lattices entering the following analysis have values of $m_\pi L\gtrsim 4$. 
We therefore expect finite volume effects to be at the percent level.

We use M\"obius Domain Wall fermions also for the discretisation of 
charm quarks where the choice of simulation parameters
in the dynamical study is based on our findings from the 
quenched case. In particular, the choice of the Domain Wall Height turns out
to be crucial in reducing discretisation effects. We also found that discretisation
effects become sizeable abruptly as the bare input quark mass is chosen 
larger than $am_h\approx 0.45$. 
\begin{figure}
\begin{center}
	\includegraphics[width=7cm]{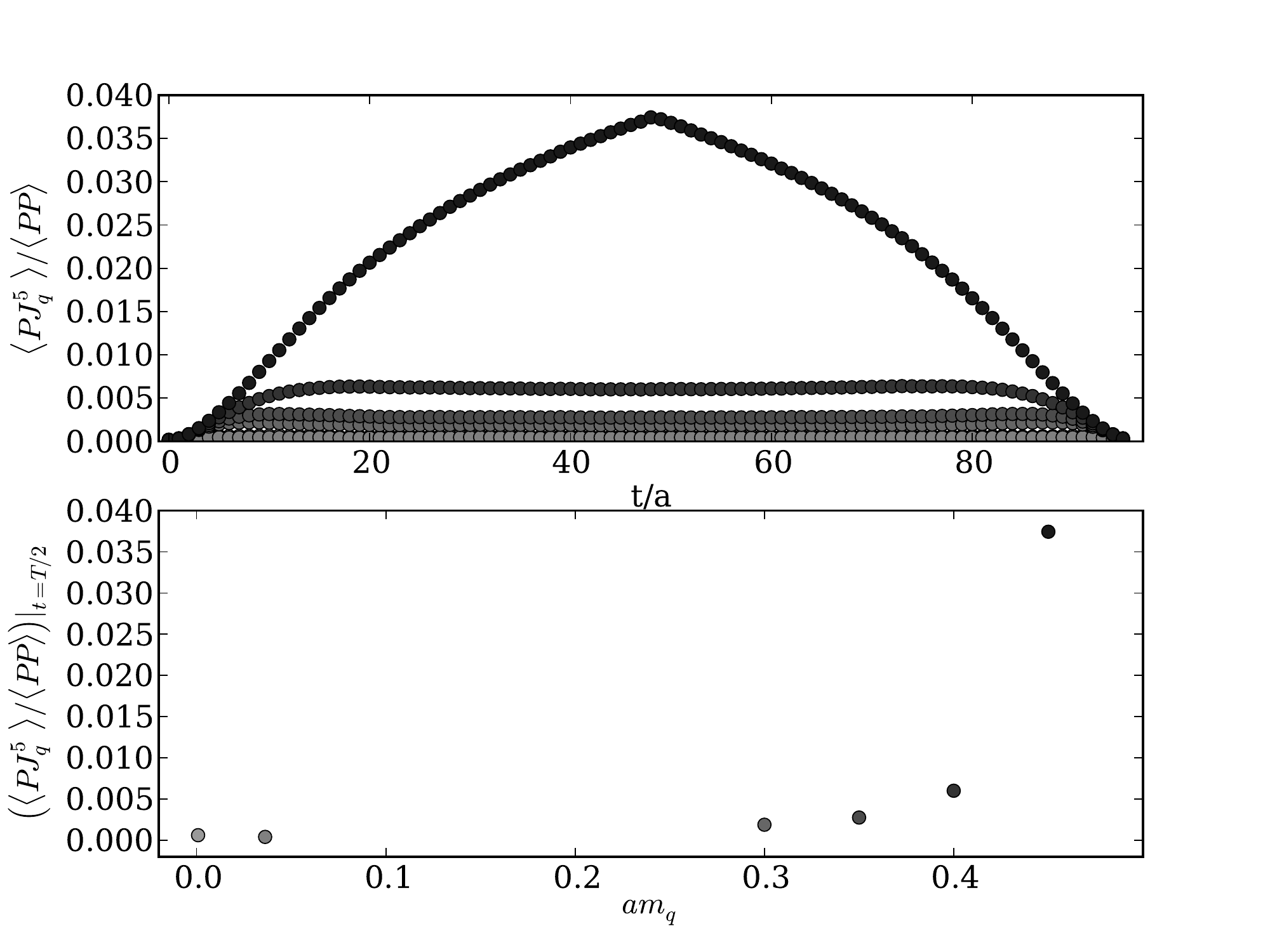}
\end{center}
\caption{Motivated by the lattice axial Ward Identity the observable in 
the top panel is constant towards the centre of the lattice for the
range of simulation parameters where the domain wall mechanism is functioning well. 
We found that the residual mass starts increasing severely as the bare input quark mass
is chosen beyond 0.45. From this point on it's no longer guaranteed
that chiral symmetry is maintained. The bottom plot shows the mass dependence 
of the residual mass (we took the value of $m_{\rm res}$ at $T/2$).}\label{fig:mres}
\end{figure}
This is also reflected in an increase in the residual mass. The residual mass
is a measure for the amount of chiral symmetry violation of domain wall fermions
due to the finite extent of the 5th dimension. The axial Ward identity for 
domain wall fermion takes the form 
\begin{equation}
\langle \partial_\mu\mathcal{A}_\mu(x)P(0)\rangle=2m\langle P(x)P(0)\rangle + 2\langle J_{5q}(x)P(0)\rangle\,.
\end{equation}
Here, $\mathcal{A}_\mu$ is the conserved axial current, $P$ is the pseudo scalar density and $J_{5q}$ is 
also a pseudo-scalar current but it mediates between the boundaries of the 5th dimension and the mid-planes.
We define the residual mass as
\begin{equation}
am_{\rm res}=\frac{\sum\limits_x \langle J_{5q}(x)P(0)\rangle}{\sum\limits_x \langle P(x)P(0)\rangle}\,.
\end{equation}
The behaviour of the residual mass as we increase the bare input quark mass is 
shown in figure~\ref{fig:mres}. 
For bare quark masses below around $am_h= 0.45$ we see a plateau which is the expected 
behaviour. For larger input quark masses the behaviour changes drastically
indicating the breakdown of the domain wall mechanism.
Based on these findings all simulations shown in the following will be for a number of 
input charm quark masses $am_h\leqq 0.45$. The heaviest charm quark we  
simulate therefore corresponds to 
$\eta_{c}$ masses of $2.8$GeV on the coarse ensemble and 3.3GeV on the medium ensemble
(for comparison the physical $\eta_{c}$ mass 2.9836(7)GeV~\cite{Agashe:2014kda}).

On the coarser physical point ensemble we generate 48 time-plane complex $Z_2$ noise
sources per configuration and on the finer ensemble 32 time-planes. Meson
twopoint functions are then computed using the one-end-trick~\cite{Foster:1998vw,Boyle:2008rh}.
On the 
unphysical point ensemble the number of source planes varies and simulations
are still ongoing. One particular concern was the convergence of the conjugate
gradient minimiser in the computation of the quark propagator. We employ the
time-slice residual~\cite{Juttner:2005ks}
\begin{equation}
r(t)=\frac {|D\psi-\eta|_t}{|\psi|_t}
\end{equation}
where $D$ is the Dirac operator, $\psi$ the solution vector and $\eta$ the 
source vector. The norm $|\cdot|_t$ is restricted to the time-slice $t$. Global
residuals will not notice convergence issues at large Euclidean time-distances
from the source in a situation where $|\psi|_t$ decays with $t$ over more orders
of magnitude than covered by the numerical precision. The time-slice residual
is a local (in time) residual that allows to control convergence more reliably.
\section{Preliminary results}
\begin{figure}
\begin{center}
 \includegraphics[width=7cm]{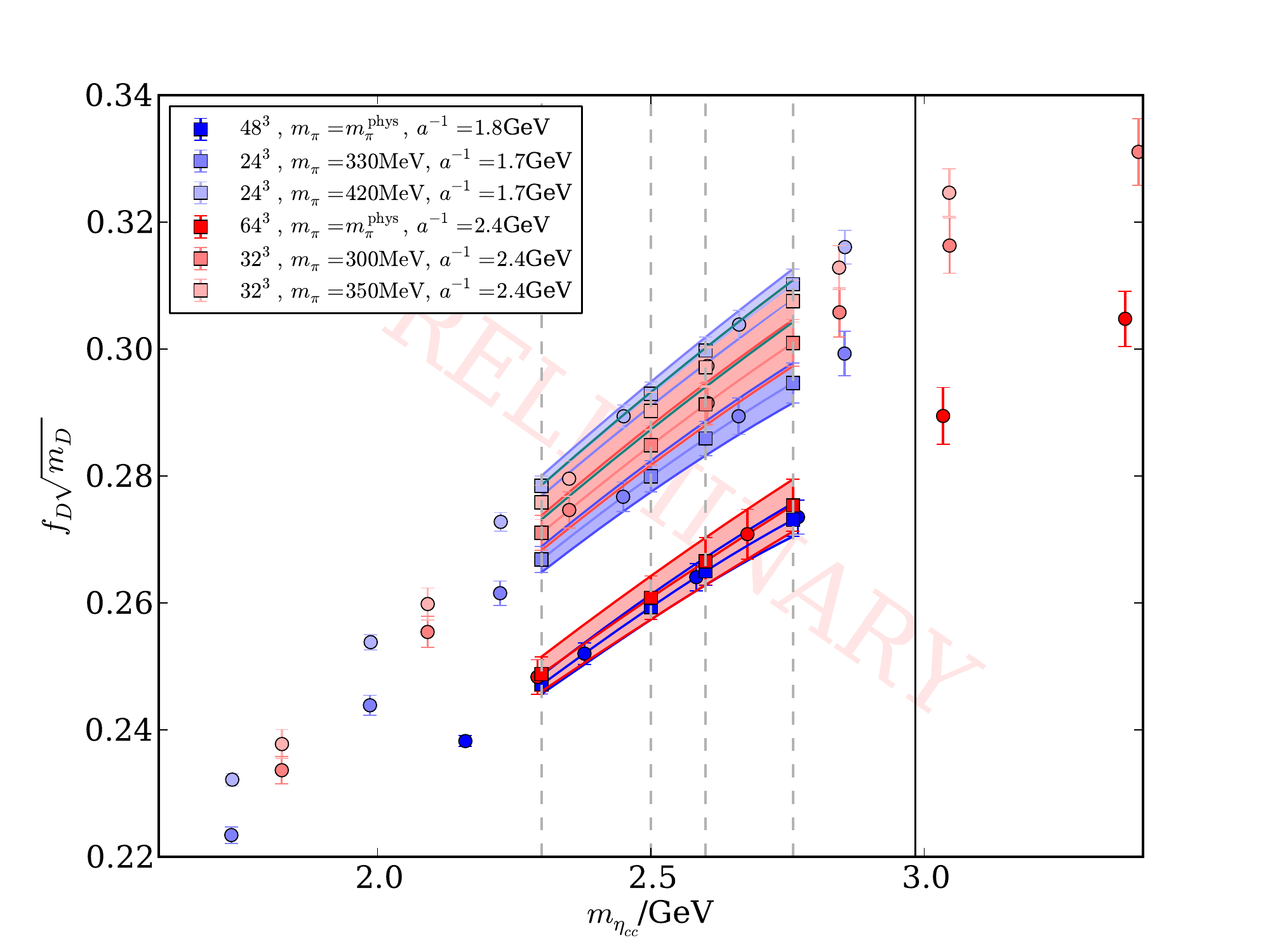}
 \includegraphics[width=7cm]{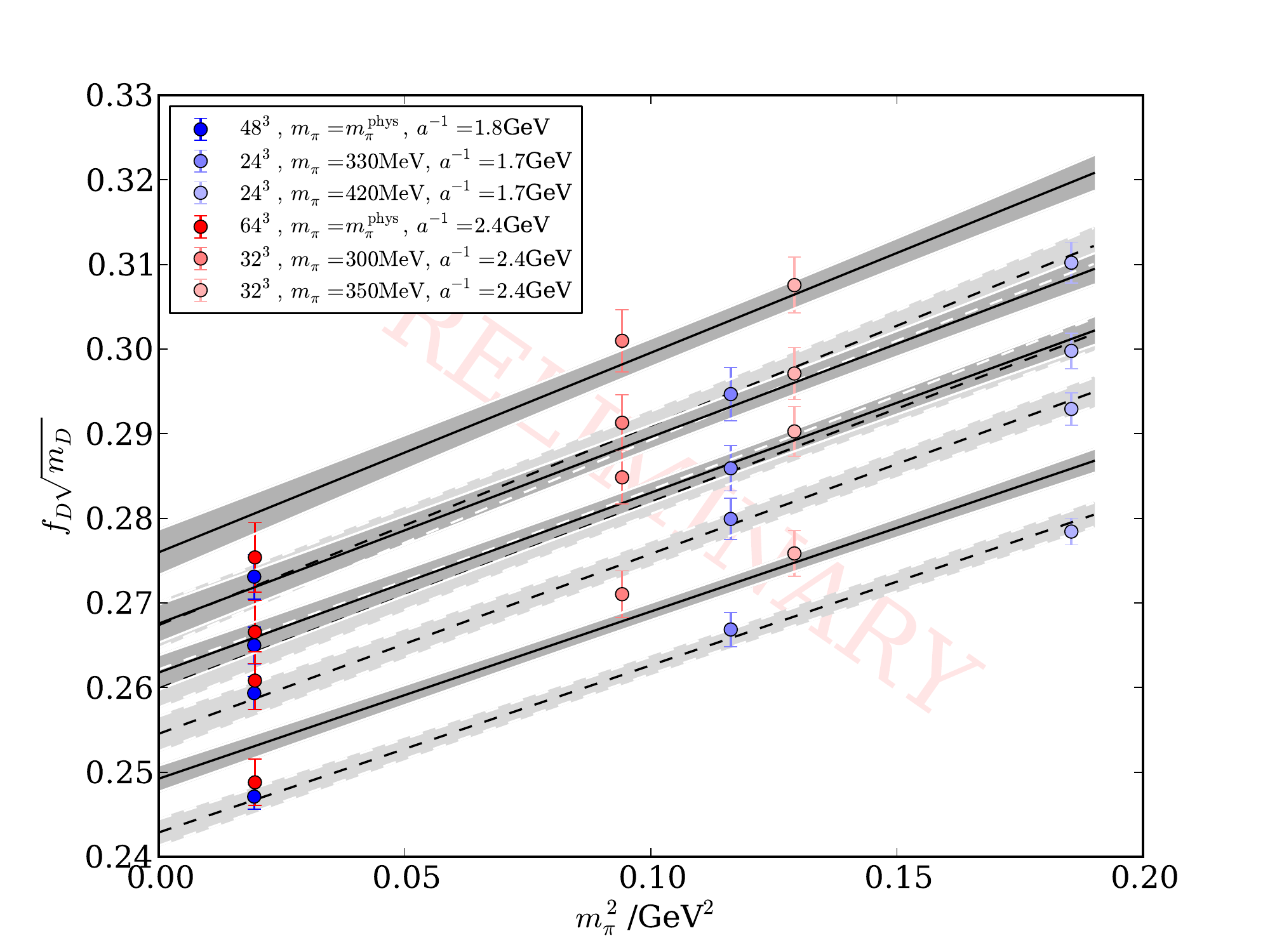}
\end{center}\caption{Left: data for the $D$-meson decay constant vs. the $\eta_c$ mass. 
The blue and red bands represent the result of a polynomial fit to the data. The
dashed vertical lines indicate the reference $\eta_c$-masses which we interpolated to.
Right: The results for the decay constant obtained for the reference $\eta_c$ masses
are then interpolated in the pion mass to its physical value. This interpolation is
over a very small range given that two of our ensembles have near-physical simulation parameters.}
\label{fig:mh and ml interpol}
\end{figure}
\begin{figure}
\begin{center}
 \includegraphics[width=7cm]{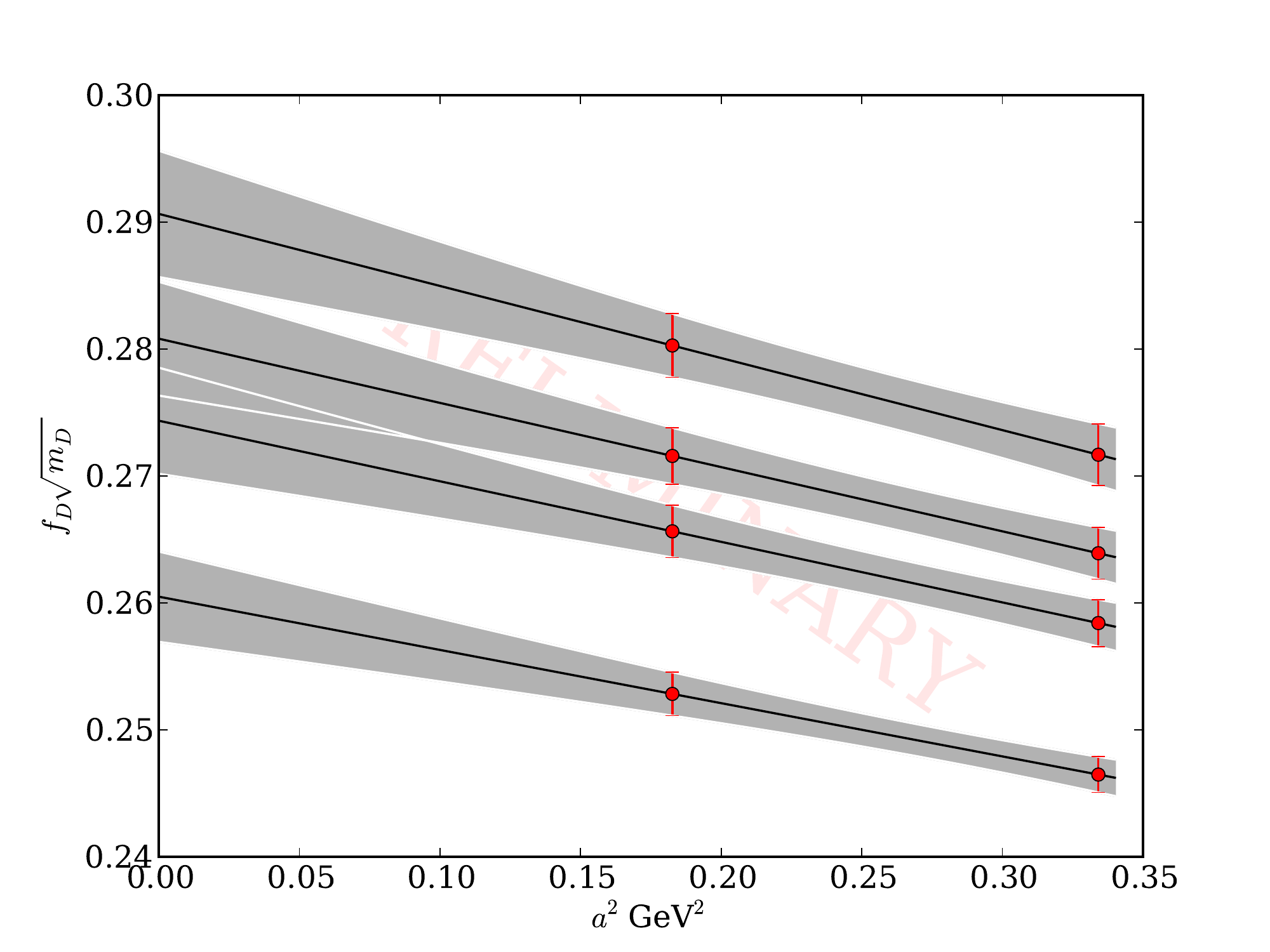}
 \includegraphics[width=7cm]{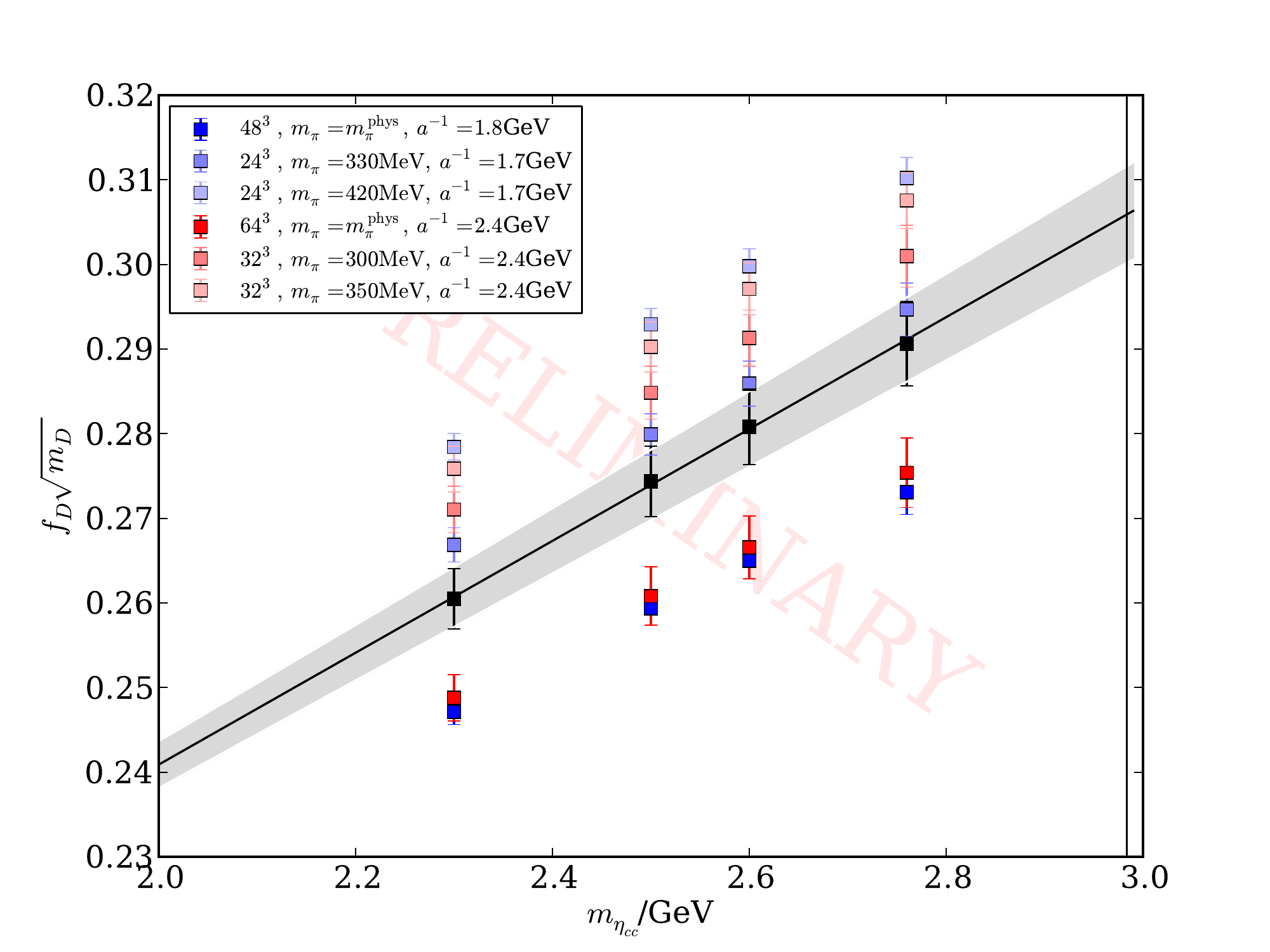}
\end{center}\caption{Left: Preliminary result for the continuum extrapolation of
the data with physical light quark mass at the reference heavy-quark mass
points. Right: Extrapolation of the data in the continuum limit (black squares)
to the physical charm point (solid vertical line to the right). }
\label{fig:cl limit and mh extrapol}
\end{figure}
The idea is to compute heavy-light and heavy-strange meson observables
for a number of unphysical heavy quark masses below and above the physical 
charm quark mass as illustrated in the l.h.s plot in figure~\ref{fig:mh and ml interpol}
for the case of the $D$-meson decay constant. 
These will then be 
interpolated to common values for the $\eta_{c}$-mass on each ensemble (the 
analysis discussed in the following uses the reference masses 
$m_{\eta_{c}}=$2.3GeV, 2.5GeV, 2.6GeV and 2.76GeV as indicated by the dashed
vertical lines). The mass dependence is close to linear and interpolating with
polynomials of various orders leads to very similar results. Results from
different orders for this interpolation can at a later stage be used to 
devise an estimate of a systematic error which will likely turn out 
to be very small.
We note that at this stage we do not
yet have full statistics and error bars for all ensembles will be much smaller
once the simulations are finalised.

The r.h.s. plot of figure~\ref{fig:mh and ml interpol} shows the results
of the above interpolation plotted against the pion mass. The points to the 
very left in this plot are from our physical point ensembles. 
Although these ensembles represent QCD with valence and sea pions close to
the physical point we still wish to study interpolation ans\"atze that allow
us to use also the results obtained on the heavier sea quark mass 
ensembles. Firstly, this will allow  for correcting small mistunings in the light
quark mass towards the actual physical point. Secondly, in the near term 
we will not be able to generate fine ensembles with physical sea quarks. 
Under the reasonable assumption that the light quark mass dependence will suffer
only very small cutoff effects our precise knowledge of the mass dependence 
on the coarse and medium ensembles will allow for extrapolating the result
on the fine ensemble towards physical light quark masses. The interpolation
shown in the r.h.s. plot in figure~\ref{fig:mh and ml interpol} are done under the
assumption that the slope with $m_\pi^2$ is independent of the mass cutoff. 
With better statistics on the coarse and medium ensemble we will soon be able 
to make more rigid statements about the validity of this assumption.

The interpolation in the light quark mass is followed by the continuum extrapolation.
The l.h.s. plot in figure~\ref{fig:cl limit and mh extrapol} can only be indicative
-- a continuum limit with only two points  for charm observables is very risky and
no quantitative phenomenological conclusions should be drawn from our data at the 
moment. This
will have to wait until we have results for the finer lattice spacing.
The r.h.s. plot in figure~\ref{fig:cl limit and mh extrapol} shows the extrapolation
of the results in the continuum limit to the physical charm point (solid vertical line)
using a linear ansatz.
\section{Conclusions and outlook}
In this talk we presented the status of our new charm physics program based on
RBC/UKQCD's $N_f=2+1$ domain wall fermion ensembles with an added valence
domain  domain wall charm quark. We have shown first results for the
$D$-meson decay constant. As anticipated based on the data from our 
exploratory quenched study we expect rather mild cutoff effects for our
choice of simulation data and judging from the data on our coarse
and medium lattice this remains so also in the dynamical case. 

We are currently generating an ensemble for a third lattice spacing which
will allow for reliably extrapolating our results to the continuum limit.
Together with the data for  the leptonic $D$- and $D_s$
decay constant we are also generating heavy-heavy correlators and 
also the correlation functions relevant for determining the  meson
bag parameter and semi-leptonic meson decays. 
Further to charm phenomenology we are studying how to make use of results for
heavier than charm quark masses which we are generating in particular on the
medium and fine ensemble. To this end we have in mind interpolating with results
in the static limit (HQET) but also applying the \emph{ratio method}~\cite{Blossier:2009hg}.

\noindent{\bf Acknowledgements:}
The research leading to these results has received funding from the Euro\-pe\-an
Research Council under the European Union's Seventh Framework Programme
(FP7/2007-2013) / ERC Grant agreement 279757.
PAB acknowledges support from STFC grants ST/L000458/1 and ST/J000329/1 and NG acknowledges the STFC grant ST/J000434/1.
The authors gratefully acknowledge computing time granted through the STFC funded DiRAC facility (grants ST/K005790/1, ST/K005804/1, ST/K000411/1, ST/H008845/1).

\bibliographystyle{elsevier}
\bibliography{juettner}

\begin{thebibliography}{10}

\bibitem{Aoki:2013ldr}
S.~Aoki et~al., Eur.Phys.J. C74(9) (2014) 2890, 1310.8555.

\bibitem{Tsangtalk}
A.~J{\"u}ttner et~al.  (2015), 1501.00660.

\bibitem{Chotalk}
Y.G. Cho et~al.  (2014), 1412.6206.

\bibitem{Allton:2008pn}
RBC-UKQCD Collaboration, C.~Allton et~al., Phys.Rev. D78 (2008) 114509,
  arXiv:0804.0473.

\bibitem{Aoki:2010pe}
Y.~Aoki et~al., Phys.Rev. D84 (2011) 014503, 1012.4178.

\bibitem{Aoki:2010dy}
RBC Collaboration, UKQCD Collaboration Collaboration, Y.~Aoki et~al., Phys.Rev.
  D83 (2011) 074508, 1011.0892.

\bibitem{Blum:2014tka}
RBC Collaboration, UKQCD Collaboration Collaboration, T.~Blum et~al.  (2014),
  1411.7017.

\bibitem{Iwasaki:1984cj}
Y.~Iwasaki and T.~Yoshie, Phys. Lett. B143 (1984) 449.

\bibitem{Iwasaki:1985we}
Y.~Iwasaki, Nucl. Phys. B258 (1985) 141.

\bibitem{Kaplan:1992bt}
D.B. Kaplan, Phys. Lett. B288 (1992) 342, hep-lat/9206013.

\bibitem{Shamir:1993zy}
Y.~Shamir, Nucl. Phys. B406 (1993) 90, hep-lat/9303005.

\bibitem{Brower:2004xi}
R.C. Brower, H.~Neff and K.~Orginos, Nucl.Phys.Proc.Suppl. 140 (2005) 686,
  hep-lat/0409118.

\bibitem{Brower:2005qw}
R.~Brower, H.~Neff and K.~Orginos, Nucl.Phys.Proc.Suppl. 153 (2006) 191,
  hep-lat/0511031.

\bibitem{Brower:2012vk}
R.C. Brower, H.~Neff and K.~Orginos  (2012), 1206.5214.

\bibitem{Agashe:2014kda}
Particle Data Group Collaboration, K.~Olive et~al., Chin.Phys. C38 (2014)
  090001.

\bibitem{Foster:1998vw}
UKQCD Collaboration Collaboration, M.~Foster and C.~Michael, Phys.Rev. D59
  (1999) 074503, hep-lat/9810021.

\bibitem{Boyle:2008rh}
P.~Boyle, A.~J{\"u}ttner, C.~Kelly and R.~Kenway, JHEP 0808 (2008) 086,
  0804.1501.

\bibitem{Juttner:2005ks}
A.~J{\"u}ttner and M.~Della~Morte, PoS LAT2005 (2006) 204, hep-lat/0508023.

\bibitem{Blossier:2009hg}
ETM Collaboration Collaboration, B.~Blossier et~al., JHEP 1004 (2010) 049,
  0909.3187.

\end{thebibliography}
%

\end{document}